\documentclass[aip,reprint]{revtex4-1}
\usepackage{graphicx}
\usepackage{svg}
\usepackage{amsmath}
\usepackage{soul}
\usepackage{xcolor}
\sethlcolor{yellow}
\usepackage{comment}
\usepackage{booktabs}
\usepackage{siunitx}
\draft 

\newcommand{\ket}[1]{|\,#1\rangle}

\begin{document}

\title{Dual wavelength source of entanglement for space quantum communication}

\author{Valentin Dumas}
\affiliation{Université Côte d’Azur, CNRS, Institut de Physique de Nice, 06108 Nice Cedex 2, France}
\author{Alek Lagarrigue}
\affiliation{Université Côte d’Azur, CNRS, Institut de Physique de Nice, 06108 Nice Cedex 2, France}
\author{Tess Troisi}
\affiliation{Université Côte d’Azur, CNRS, Institut de Physique de Nice, 06108 Nice Cedex 2, France}
\author{Gregory Sauder}
\affiliation{Université Côte d’Azur, CNRS, Institut de Physique de Nice, 06108 Nice Cedex 2, France}
\author{Sébastien Tanzilli}
\affiliation{Université Côte d’Azur, CNRS, Institut de Physique de Nice, 06108 Nice Cedex 2, France}
\author{Anthony Martin}
\affiliation{Université Côte d’Azur, CNRS, Institut de Physique de Nice, 06108 Nice Cedex 2, France}
\author{Olivier Alibart}
\affiliation{Université Côte d’Azur, CNRS, Institut de Physique de Nice, 06108 Nice Cedex 2, France}
\date{\today}

\begin{abstract}
We report the demonstration of a bulk, intrinsically phase-stable source of polarization- and time-energy-entangled photon pairs at 810\,nm and 1550\,nm, directly coupled into single-mode optical fibers. This highly non-degenerate wavelength combination is well suited for hybrid quantum communication networks, enabling low-loss transmission in optical fibers at 1550\,nm while maintaining efficient free-space propagation and detection at 810\,nm. The source is based on spontaneous parametric down-conversion in a periodically poled lithium niobate crystal embedded in a polarization Sagnac interferometer, providing inherent stability and dual-degree-of-freedom entanglement.
We measure a spectral brightness of $B = 4.8\times10^3$ pairs$\cdot$s$^{-1}$mW$^{-1}$GHz$^{-1}$, with fiber coupling efficiencies exceeding 48\% at both wavelengths. The entanglement quality is characterized by high-visibility two-photon interference, yielding net visibilities of $99.5\%$ in the polarization basis and $99.1\%$ in the energy–time basis. These performances demonstrate a compact and robust entanglement source compatible with hybrid fiber/free-space quantum key distribution architectures, and suitable for future ground-to-satellite quantum communication links.

\end{abstract}


\maketitle 

\section{Introduction}

Entanglement is a cornerstone of quantum technologies with no classical counterpart, and its distribution is essential for applications including entanglement-based quantum key distribution, networked quantum computing, and distributed quantum sensing. Currently, key objectives include the implementation of quantum systems for real-world applications, as well as the investigation of practical configurations for solving a major bottle-neck : Network connections are limited to $\sim$200\,km by fiber losses or Earth’s curvature in the case of free-space optical communication \cite{Ursin2007_Entanglement144km}.
Space-based quantum communication provides a viable solution, offering the potential to connect remote metropolitan quantum key distribution (QKD) networks via satellite links\cite{SpaceQKD-chinois}\textsuperscript{,}\cite{deForgesdeParny2023}. Such systems exploit the lower photon losses in free-space propagation in the atmosphere, enabling secure key distribution over thousands of kilometers. In this context, the demonstration of versatile, efficient and reliable sources of photonic entanglement is a stepping stone for further exploration of practical infrastructure topology. In particular, exploring the gap between space-based QKD and fibered QKD networks requires entanglement sources optimized for both free space and fiber propagation, which implies specific considerations in terms of wavelengths and entangled observables.\\


We report here a high performance source generating highly non-degenerate photons offering simultaneously energy-time and polarization entanglement that would potentially serve as a quantum interconnect between varied networks comprising space, airborne, and terrestrial assets, or, more generally, any two networks of differing wavelengths (i.e. involving a solid-state qubit). The strategic choice of wavelengths of 1550\,nm and 810\,nm are justified by their complementary roles: on the one hand, the infrared photons are intended for terrestrial fiber network, to exploit the low losses and mature infrastructure of telecom optical fibers. On the other hand, free-space optical links to satellites will benefit from the shorter wavelength photons at 810\,nm, which offers reduced beam divergence and practical detection using compact and space compatible Silicon avalanche diodes optimized for visible and near-infrared wavelengths. For extended versatility, the device offers energy-time entanglement naturally by spontaneous parametric down-conversion but also polarization entanglement thanks to a polarization dependent interferometric setup. An important aspect of the work is the injection in single mode fibers guaranteeing the produced photon pairs to be in a single (spatial) mode, thus increasing the correlation quality and practical use in real field scenario.

To this end, we took advantage of a nonlinear crystal nested within a Sagnac interferometer for its inherent stability, minimizing phase drift and optimized simultaneously the heralding efficiency into single mode optical fibers and the generation rate. In the paper, we will describe the setup of the source and characterize its main figures of merit, i.e. brightness, coupling efficiency and entanglement quality over both observables.

\section{Experimental set-up}

\begin{figure}[htb]
\centering\includegraphics[width=1.03\columnwidth]{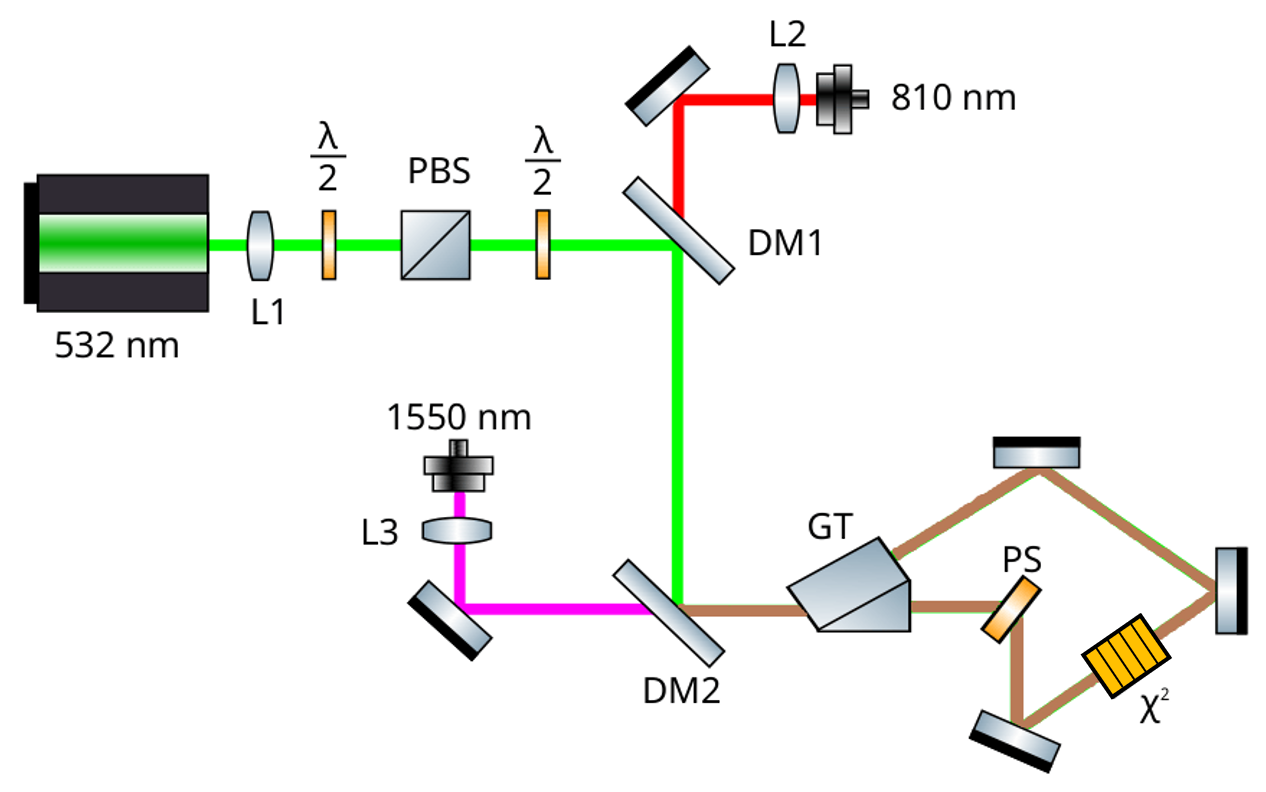}
\caption{Schematic of the highly non-degenerate entangled photon pair source. The polarization dependent Sagnac configuration enables to exploit both energy-time and polarization entanglement by exploiting the two counter propagating modes within the Sagnac loop. The periscope inside the interferometer (PS) applies a perfect $90^\circ$ polarization rotation on the three interacting wavelengths. The Glan-Thompson polarizer (GT) also acts as an achromatic polarizing beamsplitter defining the two counter-propagating modes focused into the periodically poled lithium niobate (PPLN). Further chromatic elements are used on individual modes such as two dichroic mirrors with cutoff respectively at 735\,nm (DM1) and 925\,nm (DM2) for the separation of the photon pairs. A set of half-wave plate ($\frac{\lambda}{2}$) and polarizing beam-splitter (PBS) at 532\,nm are used for power control.}
\label{sagnac_schema}
\end{figure}

As shown in \figurename~\ref{sagnac_schema}, the pump laser is a 532\,nm single frequency laser (Laser Quantum, Torus) whose polarization and spatial mode is cleaned by single mode PM fiber, such as the output pump from the fiber is linearly polarized. A low-pass filter is added to remove any residual 1064\,nm photons from the Nd:YAG inside the pump laser. A set of half-wave plate and quarter-wave plate control the polarization state of the pump photons to be $\frac{\ket{H}+e^{i\phi_p}\ket{V}}{\sqrt{2}}$.
A doublet lens ($f = 7.5$ mm) is used to focus the light from the fiber to the middle of the periodically poled lithium niobate doped MgO 5\%, type-0 and 10 mm long (Covesion, LSFG1-0.5-10). The phase matching for the SPDC 532\,nm $\rightarrow$ 810\,nm +1550\,nm, is obtained at the temperature of $T=80.10^{\circ}C$.
This crystal is nested in an achromatic Sagnac loop composed of a Glan-Thompson (GT) polarizer separating the pump beam along two paths clockwise (CW) and counter-clockwise (CCW), associated to  individual polarization mode H and V respectively.
The polarization of both paths is flipped thanks to a double periscope  acting as an achromatic half-wave plate on the pump and SPDC photons~\cite{Hentschel:09}. Both $\ket{HH}$ and $\ket{VV}$ pairs amplitudes recombine coherently on the GT polarizer. Finally, a set of two dichroic mirrors (DM1 and DM2 from Semrock) are used to separate the pump from the signal and idler photons, with an extinction ratio of 30\,dB, before their injection into single mode fibers at their respective wavelength. Here a careful calculation of the pump waist as well as the collection waists were chosen following the optimization method presented in reference~\cite{Guerreiro:13}, maximizing coupling efficiency while preserving high generation rate as shown in TABLE~\ref{table:lens} for a 10\,mm-long PPLN crystal. Note that two additional 10\,nm band-pass filter with 60\,dB of extinction ratio are placed on the optical path of the 810\,nm and 1550\,nm photons to remove any residual pump photons.

Eventually, the generated quantum state available at the output of two single mode fibres exhibits polarization entanglement, arising from the Sagnac interferometric setup, and time-frequency entanglement, due to energy conservation in the SPDC process, resulting in strong time correlation and frequency anti-correlation. The resulting hyperentangled state, defined over these two degrees of freedom, is described as  $$|\psi\rangle=\int dt_sdt_i\;\mathcal{A}(t_s,t_i)|t_s,t_i\rangle\otimes\frac{1}{\sqrt{2}}\left(|H_s,H_i\rangle+|V_s,V_i\rangle\right)$$ where $\mathcal{A}(t_s,t_i)$ denotes the joint temporal amplitude of the two-photon state.
The polarization state analysis is performed using projective measurements upon polarizing beamsplitters~\cite{QT-kwiat} or violation of Bell inequalities~\cite{Aspect82} whereas the energy-time analysis is performed using a Franson type experiment~\cite{Fallon:25}.

\begin{table}[h]
\centering
\begin{tabular}{ |l||l|l|l| } 
 \hline
 Optical mode  & 532\,nm & 810\,nm & 1550\,nm\\
 \hline
Rayleigh length & 240\,mm & 80\,mm & 40\,mm\\
Optical waist in PPLN & 200\,$\mu$m & 145\,$\mu$m & 140\,$\mu$m\\
 \hline
\end{tabular}
\caption{Calculated focusing parameters for optimizing the coupling efficiency following reference~\cite{Guerreiro:13} using standard single mode fibre of reference SM450, 780HP and SMF-28.}
\label{table:lens}
\end{table}

\section{Brightness and coupling efficiency}
The overall performance of the source is assessed by characterizing four key metrics: the photon pairs spectral bandwidth, their coupling efficiency into single mode optical fibers, as well as the brightness and eventually the ratio of genuine coincidences to accidental coincidences (CAR).\\

The general setup for such a characterization is described as follow : the single photons at 810\,nm are detected with a Si-APD (Excelitas) showing an efficiency of $60\%$ and dark count rate of 200\,Hz, while the single photons at 1550\,nm  are detected using a superconducting nanowire single-photon detector (IDQuantique) showing an efficiency of $80\%$ and a dark count rate of 50\,Hz. The individual detection rates and the analysis of the coincidence rate are computed using a time-to-digital converter (IdQuantique). 
We measured an heralding efficiency of 55\% for the 810\,nm photons and of 48\% for the 1550\,nm photons generated in the PPLN in accordance with the reported performances demonstrated in~\cite{Guerreiro:13}. 

The bandwidth of the idler photons was measured with a tunable wavelength filter (EXFO, XTM-50), and estimated using a sinc$^2$ fit to be $\Delta\lambda_i=$2.3\,nm ($\Delta\nu_i =$ 290\,GHz). Additionally, the downconversion process from a continuous wave laser implies that the signal photon (810\,nm) has the same bandwidth in frequency as the idler ($\Delta\nu_i = \Delta\nu_s$), resulting in $\Delta\lambda_s =$ 0.63\,nm. The narrow bandwidths we measured comes from the highly non-degenerate phase matching relation between our pump photons and the pairs generated.\\

The brightness of our source has been measured to be: $B = 4.8\cdot 10^3$ pairs $\cdot$ s$^{-1}$mW$^{-1}$GHz$^{-1}$. When compared to the state of the art in terms of non-degenerate bulk sources of entanglement (see TABLE~\ref{table:3}), a higher brightness should be achievable by increasing the pump focusing in the crystal, but doing so would decrease the heralding efficiency~\cite{Guerreiro:13}. We chose to maximize our heralding efficiency since it is the most important for entanglement distribution.\\

\begin{figure}[htb]
        \centering
        \includegraphics[width=1.0\columnwidth]{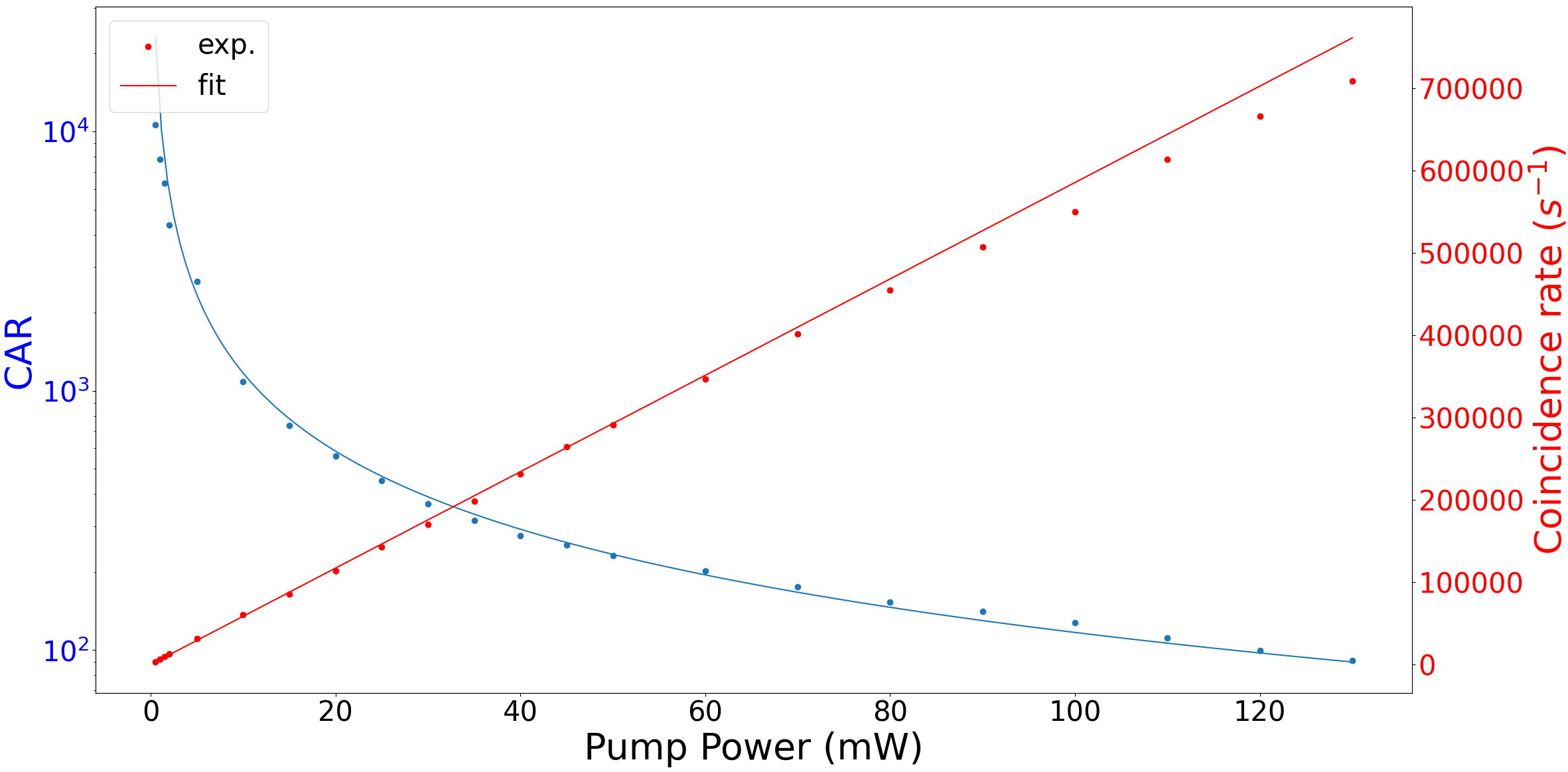}
        \caption{Evolution of the CAR and the coincidence rate in regard to the pump power. The CAR is fitted with the inverse of the pump power. This measurement is done exploiting the full range of optical power provided by our pump laser.}
        \label{CAR}
    \end{figure}

Beyond brightness, important figures of merit for practical quantum communication experiment are the coincidence rate and its associated coincidence to accidental ratio (CAR) for a given pump power. Owing to the spontaneous nature of SPDC, the photon pair statistics is poissonian. It therefore implies that two independent photon pairs can be generated simultaneously, which results in a reduced visibility when assessing quantum correlations within a time windows. The probability of emitting more than one pair per time windows increases linearly with the pump power. We reported on \figurename~\ref{CAR} the evolution of the coincidence rate for various pump power for a coincidence time windows of about 900\,ps.

The estimated mathematical fit on \figurename~\ref{CAR} indicates that when the pump laser is at full power (125\,mW), the amount of accidental coincidences represents 1.1\% of all the coincidences detected. This number can be linked to the Quantum Bit Error Rate (QBER) for an entanglement-based QKD experiment, which represents the ratio of erroneous correlations measured over the total number of correlation. In our case, pump powers beyond 1\,W would be required to increase the double-pair emissions probability to be non-negligible.

\section{Polarization and energy-time entanglement}

\begin{itemize}

    \item \textbf{Energy-time}

    As mentioned earlier, when using a highly coherent single frequency laser, the SPDC process naturally ensures energy-time entanglement. In order to reveal and characterize this specific entanglement, we perform a Franson-type experiment \cite{franson1989bell} and evaluate two-photon interference visibility.
    To do so, each photons of the pairs is sent to a pair of identical unbalanced Michelson interferometers (Exail). 
    The two interferometers shows an identical FSR of 2.5\,GHz, shorter than the bandwidth of the individual photons of about 100\,GHz, but also longer than than the bandwidth of the pump laser of about few kHz. Their optical quality has been assessed individually using classical light and they reported interference contrast of 99.9\%. Furthermore, the two interferometers show high phase stability for several hours, and their relative phase is adjustable using a piezo-mirror.
    
    We obtained a near-perfect visibility of $99.1\%$ as shown in \figurename~\ref{energy_time_vis}. These results ensure the suitability of this source for a QKD link exploiting the energy-time degree of freedom of the photons.

    \begin{figure}[htb]
        \centering
        \includegraphics[width=\columnwidth]{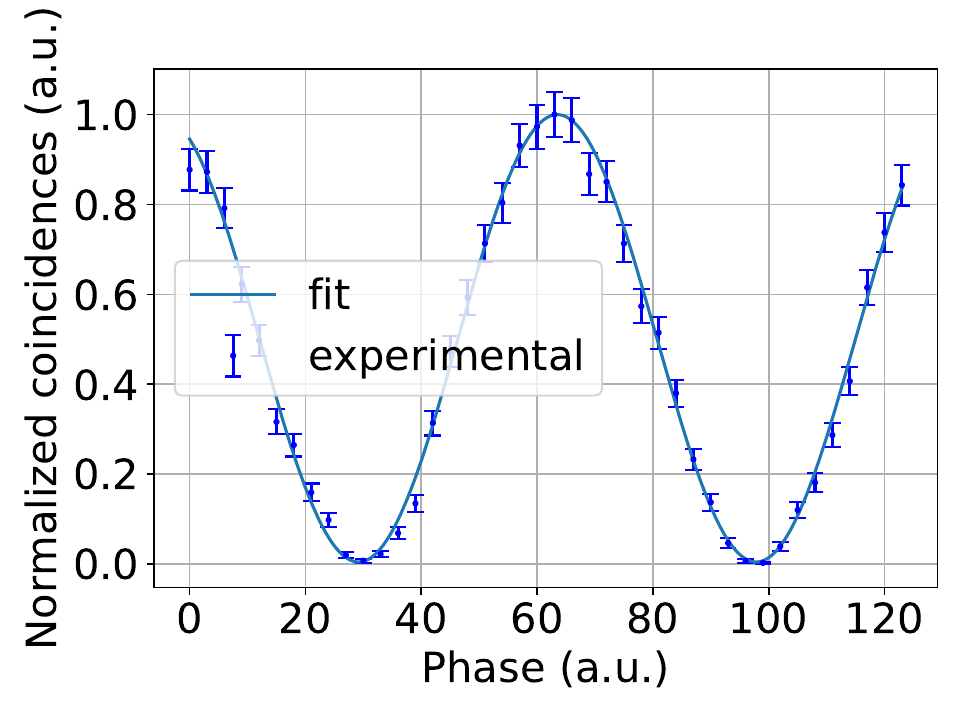}
        \caption{Energy-time entanglement fringes for the Franson-type experiment. The phase shift is expressed in Volt as we linearly controlled a piezo-electric stack mounted mirror. The measurement has been performed for a pump power of 35\,mW}
        \label{energy_time_vis}
        \label{fig:energy_time_and_vis}
    \end{figure}

    \item \textbf{Polarization}

    Quantum correlations for the polarization observable has been measured using an controllable tomography bench. It includes half-wave and quarter-wave plates associated to a polarization beamsplitter (PBS), for each idler and signal modes, allowing for the projection of each photon on different combinations of the ($H,V$), ($D,A$), and ($R,L$) polarization basis. Such a setup enables to report the entanglement quality by performing interference fringes by rotating the half-wave plate in one mode or perform a full quantum state tomography.
    
    Here, the measured interference visibility obtained for several projection scenario, enable to estimate the entanglement quality. We report on \figurename~\ref{fig:polar} the experimental measurement with their associated mathematical fit showing a net visibility of $99.5\% \pm 0.1\%$ for the H/V basis and $99.3\% \pm 0.1\%$ for D/A basis. 
    To further characterize the polarization state generated by our source, we then perform polarization quantum tomography measurements~\cite{QT-kwiat}. The polarization entangled state generated by our source was reconstructed from a set of 16 measurements where the photon pairs were analyzed in different combinations of the ($H,V$), ($D,A$), and ($R,L$) polarization basis. The density matrix extracted from these measurements is showed in \figurename~\ref{fig:polar}, the corresponding purity is of 98,1\% and the fidelity to the Bell state $\ket{\Phi^+}$ reaches 99\%.
    These results align with the state of the art in the field of entangled photon sources reported in TABLE~\ref{table:3}. The only limiting factors are the errors on the waveplates positions and the limited extinction ratio of the polarizing beamsplitter.
        
    \begin{figure}[h]
        \centering
        \includegraphics[width=0.7\columnwidth]{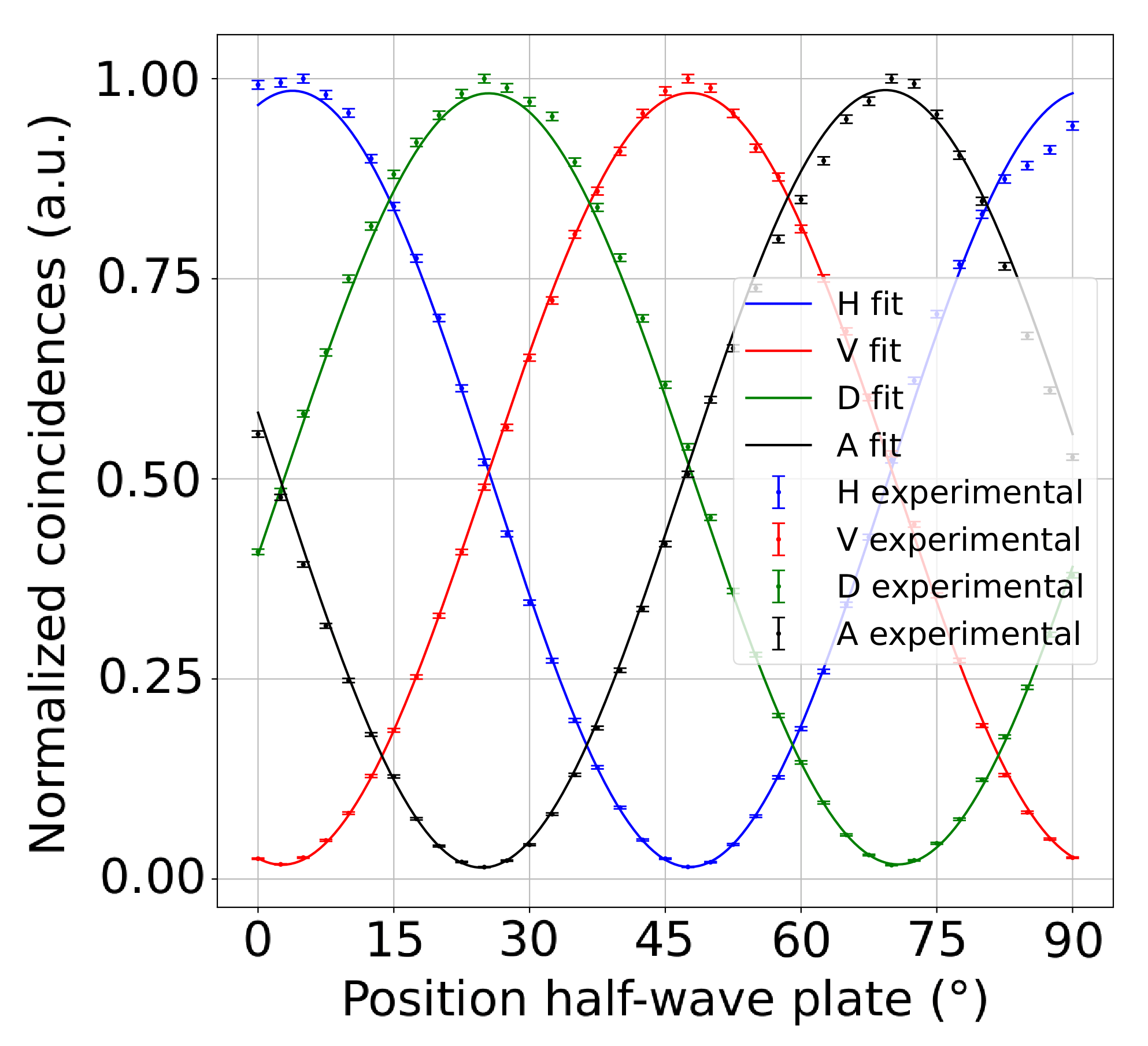}
        \includegraphics[width=1\columnwidth]{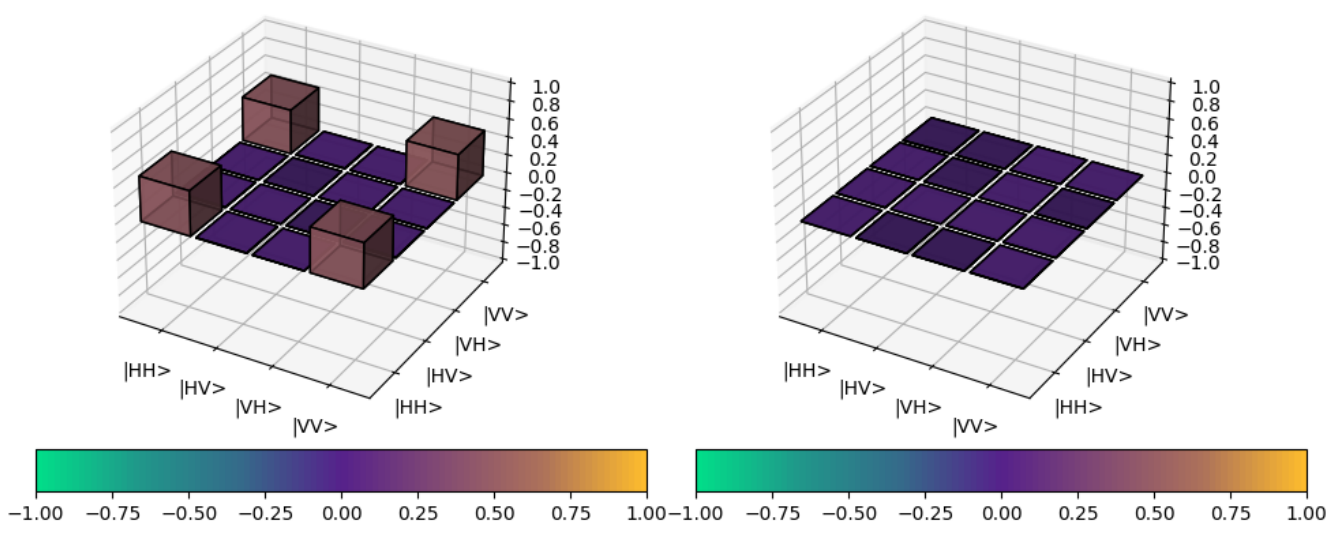}
        \caption{top : Raw polarization visibility of the bulk Sagnac source, in the horizontal (H), vertical (V), diagonal (D) and antidiagonal (A) linear polarization. The $\cos^2$ function used to fit the experimental data report a net visibility of $99.5\% \pm 0.1\%$ for the H/V basis and $99.3\% \pm 0.1\%$ for D/A basis. Bottom : Graphical representation of the real (left) and imaginary (right) parts of the density matrix extracted from experimental tomography measurements. The calculated purity is $98,1\%$ while the fidelity to the Bell state $\ket{\Phi^+}$ reaches $99\%$.}
        \label{HVDA1}
        \label{fig:polar}
    \end{figure}
    

\begin{table*}[h]
\centering
\renewcommand{\arraystretch}{1.5}
\setlength{\tabcolsep}{10pt}
\sisetup{scientific-notation=true}

\begin{tabular}{lcccc}
\toprule
\textbf{Author} & $\boldsymbol{\lambda_{s,i}}$ & \textbf{Photon Generation Rate} & \textbf{Time-Energy} & \textbf{Fidelity to $|\phi^+\rangle$} (\%) \\
& (nm)& (pairs$\cdot$ s$^{-1}$mW$^{-1}$) & \textbf{Visibility} (\%)& \\
\midrule

\textbf{This work} 
& \textbf{810, 1550} 
& \textbf{$1.4 \times 10^{5}$} 
& \textbf{99.5} 
& \textbf{99.1} \\

Peng 2025~\cite{peng2025} 
& 810, 1550 
& $8.2 \times 10^{5}$ 
& 99.69 
& 98.5 \\

Oh 2024~\cite{Oh2024} 
& 790, 1550 
& $2.9 \times 10^{6}$ 
& 97.1 
& \dots \\

Szlachetka 2023~\cite{Szlachetka2023} 
& 785, 1651 
& $6.17 \times 10^{6}$ 
& \dots 
& 96.72 \\

Hentschel 2009~\cite{Hentschel:09}  
& 810, 1550 
& $2 \times 10^{5}$ 
& \dots 
& 98.2 \\

\\

Babel 2025~\cite{Babel:25}  
& 810, 1550 
& $9.2 \times 10^{10}$ 
& \dots 
& \dots \\

Stuart 2013~\cite{Stuart2013} 
& 810, 1550 
& \dots 
& 99.1 
& \dots \\

Aizawa 2023~\cite{Aizawa23} 
& 606, 1550 
& \dots 
& \dots 
& 94.4 \\

\bottomrule
\end{tabular}

\caption{Performance table of various dual wavelength entanglement sources in the literature. We have reported the most relevant figure of merit, i.e. the photon pair generation rate per mW of pump power and the entanglement quality for two observables : Time-energy and polarisation when applicable. It is interesting to compare the first five lines whose research objectives are similar : demonstrating an entanglement source as compact and reliable as possible for quantum communication applications. The last three lines aim at different research objectives and the comparison should be interpreted as follow : On the one hand, reference~\cite{Babel:25} focused on a new waveguide fabrication technology showing the highest brightness ever while no entanglement witness has been reported. On the other hand, references~\cite{Stuart2013,Aizawa23} reported entanglement based interfaces for quantum memories whose narrow bandwidth prevents reporting meaningful brightness.  When the indicator cannot be inferred from the data we have used the \dots ~symbol.}
\label{table:3}
\end{table*}

\end{itemize}

\section{Discussion}
As mentionned earlier, our device aims at testing hybrid QKD scenario at the intersection of a free space link of 2.5\,km and one fibered link of 50\,km. Such a quantum communication testbed is already functional at the Observatoire de la Côte d'Azur, where an optical ground station is fitted with a 1.5\,m telescope for collecting 810\,nm photons. The static overall losses of this tesbed has been estimated at around 15\,dB of propagation losses in free space and 15\,dB in the fiber part.
\begin{figure}[h!]
        \centering\includegraphics[width=\columnwidth]{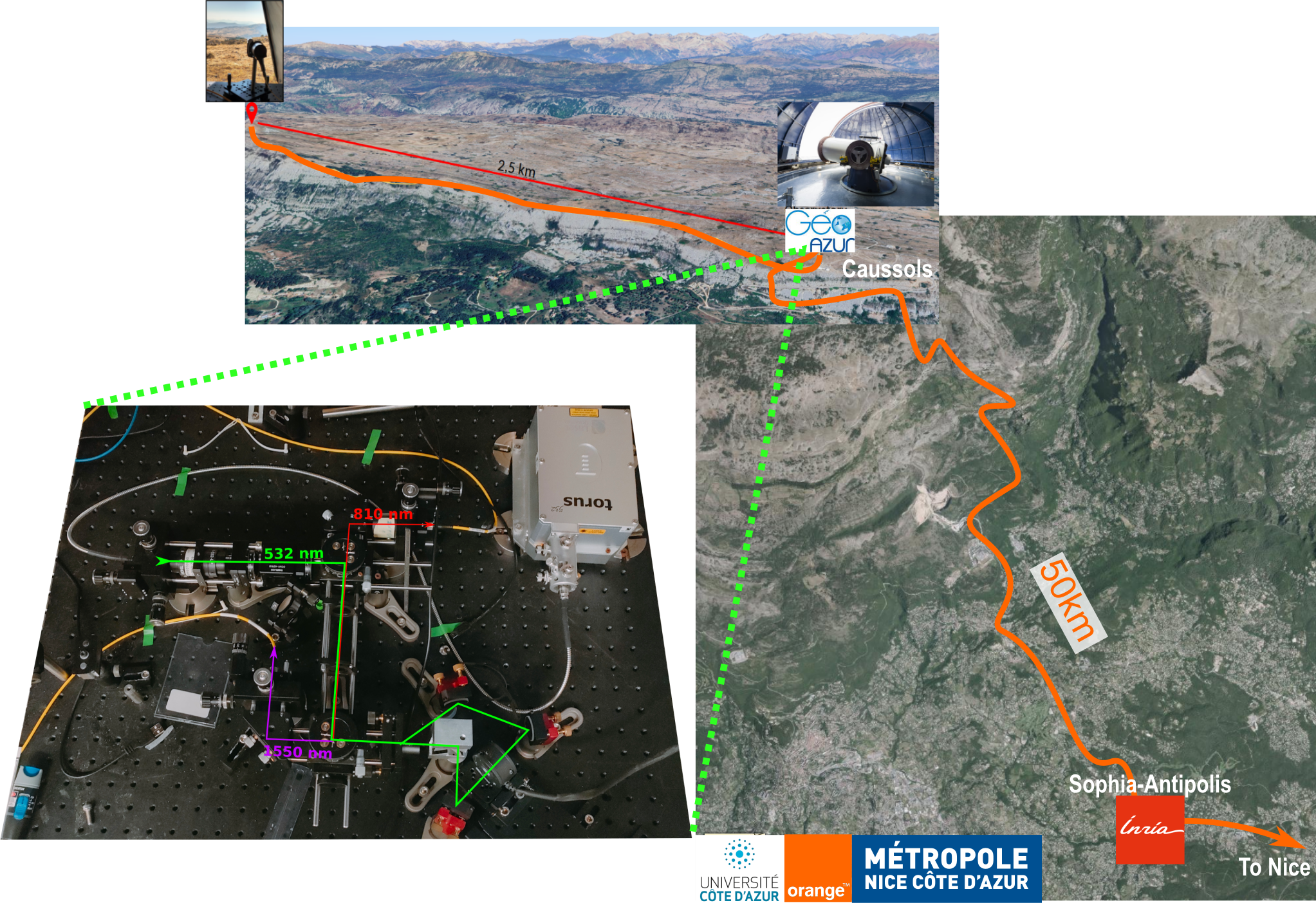}
    \caption{Hybrid QKD link with a 2.5km free space link between a reflective collimator and the Observatoire de la Côte d'azur telescope.}
    \label{qkdhybrid}
    \end{figure}
    
\begin{figure}[h!]
    \centering\includegraphics[width=1\columnwidth]{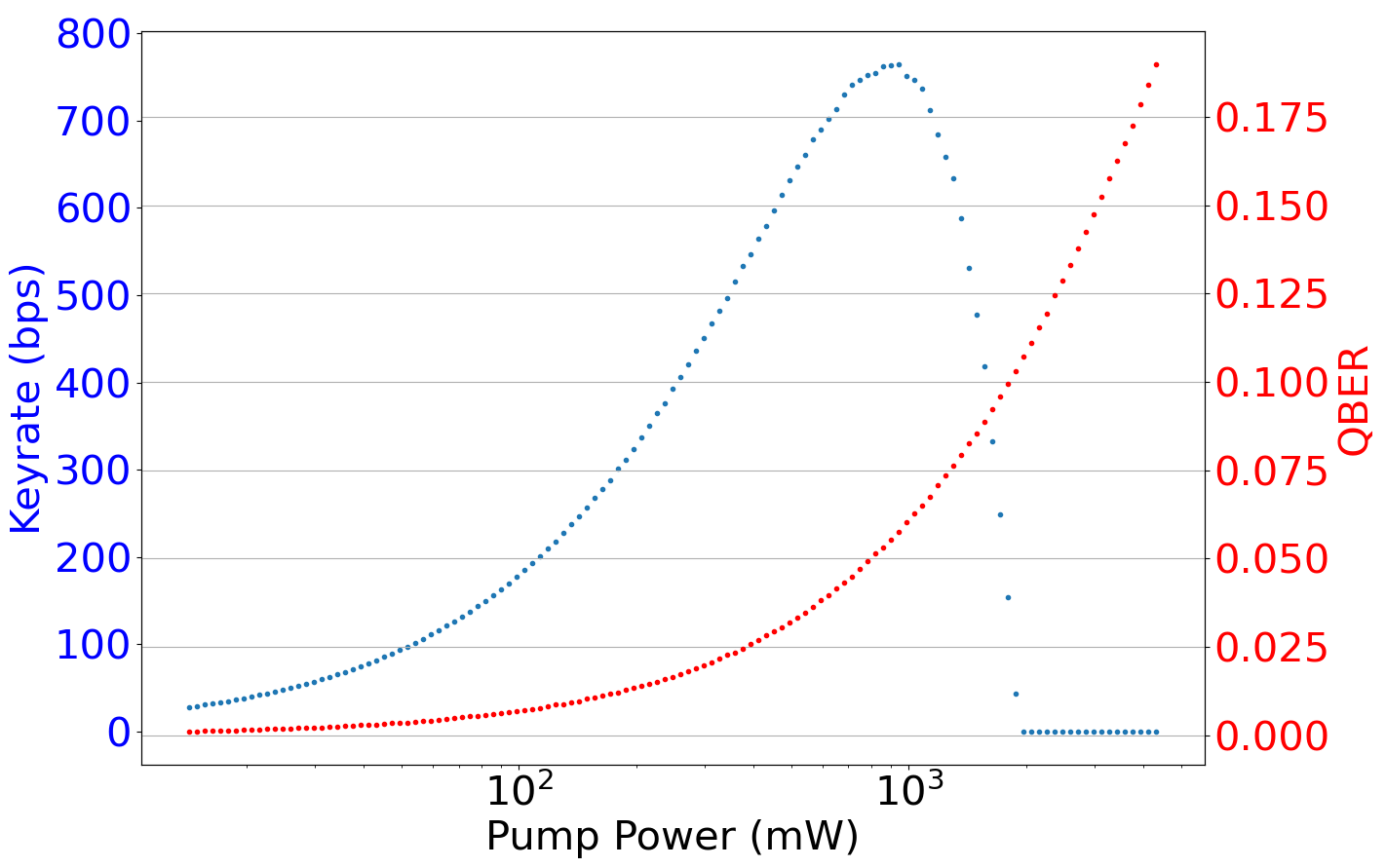}
    \caption{Simulated SKR (blue) and QBERz (red) as a function of the pump power for a hybrid link of 30\,dB losses and for a time window of 900\,ps.}
    \label{simu}
    \end{figure}

Using those estimations and the results obtained previously, we have simulated such a configuration predicted the secret key rate (SKR) and the quantum bit error rate depending on the pump power, as depicted in \figurename~\ref{simu}. While an increase of pump power enhances the rate of photon pair generation and therefore improves the SKR, it also raises the proportion of double-pair, thereby elevating the QBER and potentially reducing the overall SKR. We study both numerically the trade-off between these two effects for our loss budget, by simulating not only the quantum state of the pairs after transmission through the link and analyzers, but also by taking into account the post-treatment as described in~\cite{Cai_2009}. We found an optimal value of 5\% for the QBER in the Z basis (QBERz), which represents the maximum number of double pairs we can tolerate before the number of errors becomes too high and compromises the SNR. This study shows that, this optimal value is reached for a pump power of about $\sim 900\,mW$, well beyond our laser capabilities. However, despite not having a laser powerful enough to reach the maximum SKR, we still expect a SKR above 100\,$bps$, which would constitute an important proof of principle for space QKD, opening the door to new studies on dynamic losses and time synchronisation.

\section{Conclusion}
We demonstrated the design, performances, and characterization of a dual-wavelength entanglement source. The setup fits on a 1m$^2$ breadboard and is easy to transport. We have shown that through a careful design of the optical setup we were able to optimize simultaneously the brightness and the coupling efficiency delivering a efficient source of entangled photon pair for quantum communication experiment in real-field scenario. Additionally, we achieved a coupling efficiency of 48\% and 55\% for the 1550\,nm and 810\,nm photons respectively and we have reached a coincidence rate up to 700 kHz for 125\,mW of pump power. It is worth mentioning our relatively low brightness, due to a bulk nonlinear crystal configuration, could be advantageously improved by using an ultrabright source based on thin-film lithium niobate recently demonstrated by the group of C. Silberhorn in Germany~\cite{Babel:25}.
We have also measured a state of the art entanglement quality in both polarization and energy-time degrees of freedom enabling for practical deployment in quantum communication testbed.

Finally, our our simulations confirms our device offers promising results for the
development of scalable and robust hybrid QKD links.

\nocite{*}
\bibliographystyle{apsrev4-1}
\bibliography{aipsamp}

\begin{acknowledgments}
This work has been conducted within the framework of the French government financial support managed by the Agence Nationale de la Recherche (ANR), within its Investments for the Future program, under the Université Côte d'Azur UCA-JEDI project (ANR-15-IDEX-01), under the SOLUQs project (ANR-20-ASTQ-0003), and under the Stratégie Nationale Quantique through the PEPR-quantique projects QCOMMTESTBED (ANR 22-PETQ-0011). This work has also been conducted within the framework of the OPTIMAL project, funded by the European Union and the Conseil Régional SUD-PACA by means of the “Fonds Européens de développement regional” (FEDER). The authors also acknowledge financial support from the European Commission, through the Project 101114043-QSNP and 101091675-FranceQCI. Finally, V. Dumas acknowledges PhD funding from Metropôle Nice Côte d'Azur.
\end{acknowledgments}

\end{document}